# Reaction plane alignment with linearly polarized photon in heavy-ion collisions

Xin Wu, Xinbai Li, Zebo Tang, Pengfei Wang, and Wangmei Zha[*]
*State Key Laboratory of Particle Detection and Electronics, University of Science and Technology of China, Hefei 230026, China
and Department of Modern Physics, University of Science and Technology of China, Hefei 230026, China*

The collective observables play critical roles in probing the properties of quark-gluon-plasma created in relativistic heavy-ion collisions, in which the information on initial collision geometry is crucial. However, the initial collision geometry, e.g., the reaction plane, cannot be directly extracted in the experiment. In this paper, we demonstrate the idea of determining the reaction plane via the feature of linear polarization of the coherent photoproduction process and discuss the advantages of the proposed approach in comparison with traditional methods.

The major physical goal of relativistic heavy-ion collisions is to create quark-gluon-plasma (QGP) [1] in the laboratory and to study its properties. After decades of experimental and theoretical investigations, it is believed that the hottest manmade form of matter, QGP, has been created and discovered in the laboratory [2], reaching a milestone in the field of high-energy nuclear physics. In the next stage, the core scientific mission of relativistic heavy-ion collision physics is to study the properties of QGP and to understand the phase structure of strongly interacting quark matter [3].

Among the probes to detect the properties of QGP, the relevant observables for collective motion (e.g., anisotropic flow, global polarization, chiral magnetic effect) in heavy-ion collisions play essential roles [4–6]. Over the past decade, the collective measurements have been widely performed by the STAR, PHENIX, ALICE, ATLAS, and CMS collaborations at different collision energies and systems. The experimental results clearly demonstrate that the QGP formed in high energy A+A collisions behaves as a nearly-perfect fluid with strong coupling [7], vast vorticity [8], ultrastrong magnetic field [9], and possible chiral anomaly [10]. In these measurements, the information on the reaction plane is an indispensable prerequisite. However, at the current stage, the initial collision geometry cannot be directly extracted in the experiment. For the secondbest, the collective measurements are only estimated via the anisotropy of final particles in momentum space in traditional methods. They lose the direct connection to initial states and introduce uncorrectable bias such as nonflow correlations and event-by-event fluctuations [11], which can only be investigated in phenomenal models—an indirect way. The lack of direct information on the collision geometry weakens the sensitivities of collective measurements to quantitatively probe the properties of QGP. Therefore, there is a desperate need to build up a direct link between the initial geometry and final collective observables in experiments.

Recent research suggests that the coherent photoproduction process could exist in hadronic heavy-ion collisions (HHICs), which can provide additional information to infer the properties of QGP. In the photoproduction process, the electromagnetic field accompanied by the heavy nuclei moving at nearly the speed of light can be viewed as a spectrum of quasireal photons [12]. The equivalent photon could fluctuate into a quark-antiquark pair and then elastically scatter off the nucleus via exchange of Pomeron [13], emerging as a real vector meson, known as coherent photoproduction. Due to the highly Lorentz contraction, the induced electric field is almost fully perpendicular to the direction of motion of the heavy nuclei, which suggests that the quasireal photons are fully linearly polarized in the transverse plane. In the coherent photoproduction process, the produced vector meson inherits the linear polarization of photons, which leads to an asymmetric distribution of decay daughters. The direction of linear polarization is completely determined by the initial collision geometry, which offers us an opportunity to directly probe the initial collision geometry in the experiment. In this paper, we demonstrate the idea of probing the reaction plane via coherent photoproduction for the first time and estimate the resolution of the reaction plane from this approach at typical RHIC and LHC energies.

In relativistic heavy-ion collisions, the coherent vector meson photoproduction consists of two indistinguishable processes: either nucleus one emits a photon and nucleus two acts as a target, or vice versa. In these processes, the photon is scattered by nuclei via Pomeron exchange, which imposes a restriction on the production site within the two colliding nuclei. Furthermore, the emitted photons are fully linearly polarized, which has been suggested by Li *et al.* [14] and confirmed by the STAR Collaboration for the dielectron measurements [15]. The orientation of the photon polarization is determined by the electric vector of the electromagnetic field induced by the colliding ion, as demonstrated in Fig. 1. In the ideal case (ignoring the size and density distribution of nuclei), the polarization direction of the scattered photons is fully aligned with the impact parameter for coherent photoproduction in heavy-ion collisions. Under the helicity no-flip assumption [16–21], the produced vector meson inherits the linear polarization state, which leads to the preferential orientation of the decay angle along the direction of polarization. This offers us an opportunity to determine the reaction plane via the decay asymmetry of the vector meson from coherent photoproduction, where the reaction plane is spanned by the impact parameter and the beam axis. Here, we take the process of $\gamma + A \to \rho^0 + A \to \pi^+ + \pi^- + A$ to illustrate the idea. Herein, the right-handed coordinate system for vector meson decay is built up as follows. The $z$ axis is chosen to be the flight direction of the vector meson in the photon-nucleon center of mass frame. In the experiment, the direction of the $z$ axis can be approximated by that of colliding nuclei in the laboratory frame. The $y$ axis is normal to the reaction plane and the $x$ axis is given by $y \times z$. Following the derivation in Ref. [22], the decay angular distribution of the vector meson to two spinless daughters ($\rho^0 \to \pi^+ + \pi^-$) is

$$\frac{d^2N}{d\cos\theta d\phi} = \frac{3}{8\pi}\sin^2\theta[1+\cos(2\phi)], \quad (1)$$

where the decay angles $\theta$ and $\phi$ are the polar and azimuthal angles, respectively, which denotes the direction of one of the decay daughters in the vector meson rest frame. As revealed in Eq. 1, the pions tend to be emitted along the reaction plane. We can choose the plane spanned by the direction of one of the pions in the vector meson rest frame and the beam axis as the reaction plane ($\Psi_r = \phi$). The resolution of the reaction plane from this approach, defined as $R = \langle \cos(2\Psi_r) \rangle$, can be directly extracted from Eq. 1 to be 0.5.

---

[*]first@ustc.edu.cn

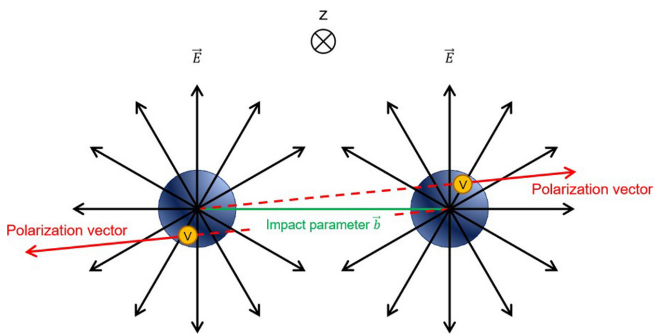

FIG. 1. Schematic diagram for the direction of the polarization direction of the photons in heavy-ion collisions. The polarization vector is radially outward along the nucleus, which emits photons.

For a realistic case, the nucleus has a finite size and specific density distribution, which would lead to a variation in the polarization direction along the impact parameter. Hereinafter, we employ the vector meson dominance (VMD) framework to estimate this variation and to reevaluate the resolution of the reaction plane from the approach described above at typical RHIC and LHC energies.

The transverse spatial photoproduction amplitude distribution for the vector meson can be expressed by convoluting the photon flux amplitude with the corresponding $\gamma A \to VA$ scattering amplitude $\Gamma_{\gamma A \to VA}$ [23]:

$$\vec{A}(\vec{x}_\perp) = (A_x, A_y) = \vec{a}(\omega, \vec{x}_\perp)\Gamma_{\gamma A \to VA}, \quad (2)$$

where $A_x$ and $A_y$ represent the production amplitude with linear polarization along the $x$ and $y$ directions, respectively.

The photon flux amplitude generated by the heavy nuclei can be given by the equivalent photon approximation (EPA):

$$\vec{a}(\omega, \vec{x}_\perp) = \sqrt{\frac{4Z^2\alpha}{\omega_\gamma}}\int \frac{d^2\vec{k}_{\gamma\perp}}{(2\pi)^2}\vec{k}_{\gamma\perp}\frac{F_\gamma(\vec{k}_\gamma)}{|\vec{k}_\gamma|^2}e^{i\vec{x}_\perp\cdot\vec{k}_{\gamma\perp}},$$

$$\vec{k}_\gamma = \left(\vec{k}_{\gamma\perp}, \frac{\omega_\gamma}{\gamma_c}\right), \quad \omega_\gamma = \frac{1}{2}M_V e^{\pm y}, \quad (3)$$

where $\vec{x}_\perp$ and $\vec{k}_{\gamma\perp}$ are two-dimensional photon position and momentum vectors in the transverse plane, $\omega_\gamma$ is the energy of the emitted photon, Z is the electric charge of the nucleus, $\alpha$ is the electromagnetic coupling constant, $\gamma_c$ is the Lorentz factor of the photon-emitting nucleus, $M_V$ and $y$ are the mass and rapidity of the vector meson, and $F_\gamma(\vec{k}_\gamma)$ is the nuclear electromagnetic form factor. The form factor can be obtained by performing a Fourier transformation to the charge density of the nucleus. We use the parameterized Woods-Saxon distribution as the nuclear charge density distribution

$$\rho_A(r) = \frac{a^0}{1+\exp[(r-R_{WS})/d]}, \quad (4)$$

where the radius $R_{WS}$ and skin depth $d$ are from the electron-scattering data [24] ($R_{WS} = 6.38$ fm, $d = 0.535$ fm for Au; $R_{WS} = 6.62$ fm, $d = 0.546$ fm for Pb), and $a^0$ is the normalization factor. The polarization direction for the quasireal photons follows the position vector $\vec{x}_\perp$ in Eq.(3).

The scattering amplitude $\Gamma_{\gamma A \to VA}$ with the shadowing effect can be obtained by the Glauber [25] plus vector meson dominance (VMD) [26] approach:

$$\Gamma_{\gamma A \to VA}(\vec{x}_\perp) = \frac{f_{\gamma N \to VN}(0)}{\sigma_{VN}}2\left[1 - \exp\left(-\frac{\sigma_{VN}}{2}T'(\vec{x}_\perp)\right)\right], \quad (5)$$

where $f_{\gamma N \to VN}(0)$ is the forward-scattering amplitude for $\gamma + N \to V + N$ and $\sigma_{VN}$ is the total $VN$ cross section. $T'(\vec{x}_\perp)$ is the modified thickness function accounting for the coherence length effect:

$$T'(\vec{x}_\perp) = \int_{-\infty}^{+\infty}\rho(\sqrt{\vec{x}_\perp^2 + z^2})e^{iq_L z}dz, \quad q_L = \frac{M_V e^y}{2\gamma_c}, \quad (6)$$

where $q_L$ is the longitudinal momentum transfer required to produce a real vector meson. The $f_{\gamma N \to VN}(0)$ can be determined from the measurements of the forward-scattering cross section $\frac{d\sigma_{\gamma N \to VN}}{dt}|_{t=0}$, which is well parametrized in Ref. [27]. Using the optical theorem and VMD relation, the total cross section for VN scattering is given by

$$\sigma_{VN} = \frac{f_V}{4\sqrt{\alpha}C}f_{\gamma N \to VN}, \quad (7)$$

where $f_V$ is the $V$-photon coupling and $C$ is a correction factor for the off-diagonal diffractive interaction [28].

In HHICs, the observation effect should be considered. For the coherent scattering process, the spectator nucleons are free from the hadronic interactions and can still act coherently. However, for the participating nucleons, the state would be affected by the violent hadronic interactions, leading to the destruction of coherent action. By taking this into account, the production amplitude is modified as

$$\vec{A}(\vec{x}_\perp) = \vec{a}(\omega, \vec{x}_\perp)\Gamma_{\gamma A \to VA} P_{noH}(\vec{x}_\perp). \quad (8)$$

$P_{noH}(\vec{x}_\perp)$ is the probability that the nucleon with position $\vec{x}_\perp$ would not suffer any hadronic interaction, given by

$$P_{noH}(\vec{x}_\perp) = (1 - T(\vec{x}_\perp)\sigma_{NN})^A, \quad (9)$$

where $\sigma_{NN}$ is the total nucleon-nucleon cross section and A is the nuclear number.

In the experiment, the measurements are performed in momentum representation. The production amplitude in momentum representation can be obtained by performing a Fourier transformation to the amplitude in coordinate representation:

$$\vec{A}(\vec{p}_\perp) = \frac{1}{2\pi}\int d^2 x_\perp (\vec{A}_1(\vec{x}_\perp) + \vec{A}_2(\vec{x}_\perp))e^{i\vec{p}_\perp \cdot \vec{x}_\perp}, \quad (10)$$

where $\vec{A}_1(x_\perp)$ and $\vec{A}_2(x_\perp)$ are the spatial amplitude distributions in the transverse plane for the two colliding nuclei.

For the coherent photoproduction at midrapidity ($y = 0$), there exists a symmetry for the spatial amplitude distribution: $\vec{A}_1(\vec{x}_\perp) = -\vec{A}_2(-\vec{x}_\perp)$. In this case, the produced vector meson in momentum representation is still linearly polarized with variations of polarization direction along the impact parameter. The decay angular distribution for the process $\rho^0 \to \pi^+ + \pi^-$ in Eq. (1) can then be modified as

$$\frac{d^2N}{d\cos\theta d\phi} = \frac{3}{8\pi}\sin^2\theta[1 + P_\gamma \cos(2\phi)]. \quad (11)$$

$P_\gamma$ is the degree of polarization along the impact parameter, which can be written as

$$P_\gamma = \left\langle \frac{A_x^2 - A_y^2}{A_x^2 + A_y^2}\right\rangle. \quad (12)$$

The equation form of $P_\gamma$ is analogous to the definition of eccentricity. The resolution of the reaction plane determination from the proposed approach can be naturally extracted as $R = P_\gamma/2$.

Figure 2 shows the calculated resolution of the reaction plane determination $R$ as a function of the impact parameter (b) for the coherent process of $\gamma + A \to \rho^0 + A \to \pi^+ + \pi^- + A$ in Au+Au collisions at $\sqrt{s_{NN}} = 200$ GeV and Pb+Pb collisions at $\sqrt{s_{NN}} = 2.76$ TeV at midrapidity for $p_T < 0.1$ GeV/$c$. As expected, the resolution of the reaction plane from this approach worsens toward central collisions. For head-on collisions ($b = 0$), there is no preferred linear polarization direction for the coherent photoproduction process, in which the proposed approach to determine the reaction plane would completely fail. In peripheral or ultraperipheral collisions, the bias of the polarization direction from nuclear size and density distribution would be small, and the resolution can approach the ideal limit ($R \to 0.5$). The resolution of the reaction plane determination is slightly better at LHC than at RHIC, which is mainly due to a more uniform photon flux over nuclei at LHC energy than that at RHIC energy. As one can find in the figure, the resolution from the proposed method is even better than that from the traditional approach ($\sim 0.3$) [29] in peripheral heavy-ion collisions. Furthermore, it provides a unique method to determine the reaction plane in ultraperipheral collisions with good resolution.

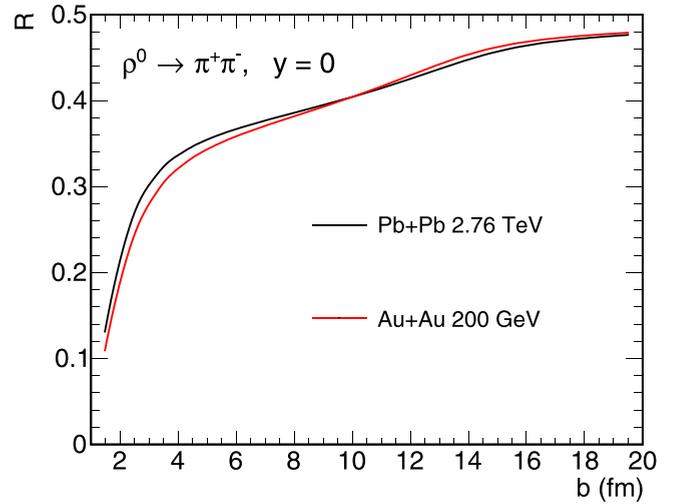

FIG. 2. The estimated resolution of the reaction plane determination $R$ as a function of the impact parameter for the coherent process of $\gamma + A \to \rho^0 + A \to \pi^+ + \pi^- + A$ in Au+Au collisions at $\sqrt{s_{NN}} = 200$ GeV and Pb+Pb collisions at $\sqrt{s_{NN}} = 2.76$ TeV at midrapidity for $p_T < 0.1$ GeV/$c$.

As revealed in Eqs. (10) and (12), the resolution of the reaction plane should depend on the transverse momentum of the coherent produced vector meson. Figure 3 shows the estimated resolution $R$ from $\rho^0$ photoproduction as a function of transverse momentum in Au+Au collisions at $\sqrt{s_{NN}} = 200$ GeV and Pb+Pb collisions at $\sqrt{s_{NN}} = 2.76$ TeV at midrapidity for $b = 10$ fm. The resolution remains almost unchanged in the low $p_T$ region ($p_T < 0.05$ GeV), and the majority of vector mesons from coherent processes are produced in this transverse momentum region. At a relatively large $p_T$, the resolution becomes worse, which is mainly because the production of vector mesons with larger $p_T$ is more sensitive to the nuclear size and density distribution. The shape difference between Au+Au collisions and Pb+Pb collisions in Fig. 3 originates from the difference in density distributions between Au and Pb.

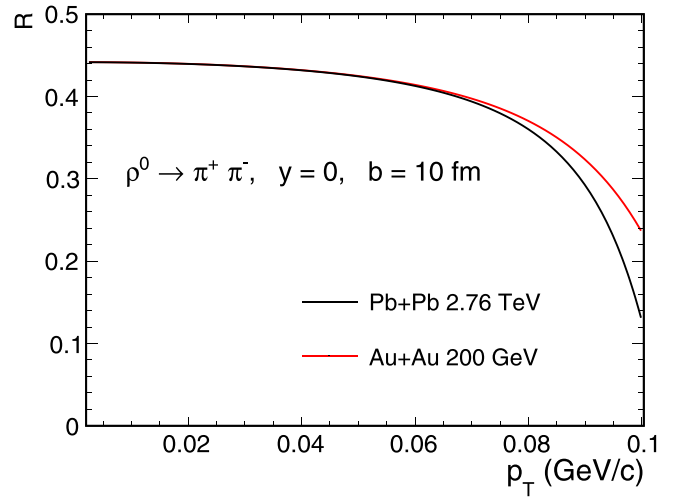

FIG. 3. The calculated resolution $R$ from $\rho^0$ photoproduction as a function of transverse momentum in Au+Au collisions at $\sqrt{s_{NN}} = 200$ GeV and Pb+Pb collisions at $\sqrt{s_{NN}} = 2.76$ TeV at midrapidity for $b = 10$ fm.

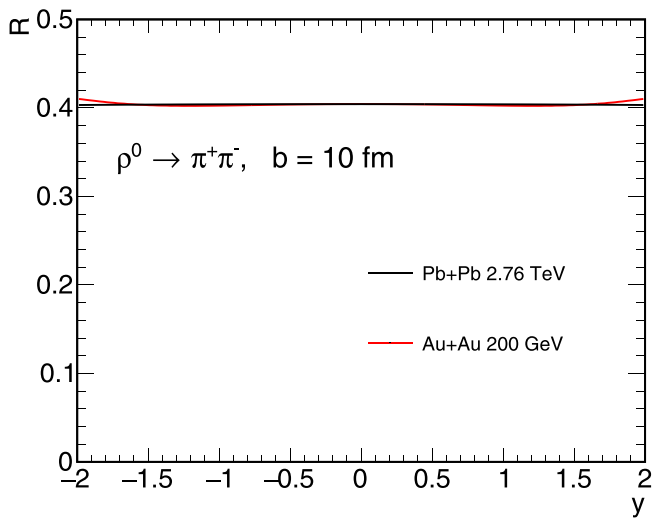

FIG. 4. The resolution $R$ as a function of the rapidity of the coherent produced $\rho^0$ for Au+Au collisions at $\sqrt{s_{NN}} = 200$ GeV and Pb+Pb collisions at $\sqrt{s_{NN}} = 2.76$ TeV with $b = 10$ fm for $p_T < 0.1$ GeV/$c$.

For the coherent photoproduction at arbitrary rapidity, the symmetry of the amplitude distribution in coordinate representation does not hold for most situations. This would lead to elliptical polarization for the production in momentum representation rather than completely linear polarization. The decay angular distribution can still be described by Eq. (11), however, the degree of linear polarization $P_\gamma$ should be rewritten as

$$P_\gamma = \left\langle \frac{A_x^2 - A_y^2}{\sqrt{(A_x^2 + A_y^2)^2 - (2\mathrm{Re}A_x\mathrm{Im}A_y - 2\mathrm{Im}A_x\mathrm{Re}A_y)^2}} \right\rangle. \quad (13)$$

For the case at $y = 0$, $A_x$ and $A_y$ are purely imaginary, which simplifies the formula to Eq. (12). The rapidity dependence of the resolution $R$ is calculated as shown in Fig. 4 for Au+Au collisions at $\sqrt{s_{NN}} = 200$ GeV and Pb+Pb collisions at $\sqrt{s_{NN}} = 2.76$ TeV with $b = 10$ fm for $p_T < 0.1$ GeV/$c$. The resolution of the reaction plane from the proposed approach is independent of the rapidity of the coherent produced $\rho^0$, and the rapidity dependence is insensitive to the collision energy and system.

Traditionally, the collective measurements are estimated via the anisotropy of final particles in momentum space in experiments, and the initial states are usually modeled via the billiard ball interaction picture encapsulated in Monte Carlo Glauber calculations [25]. The link between the initial geometry and final collective motions is built up indirectly by the transport and hydrodynamic models [30–32]. Furthermore, the uncorrectable nonflow correlations and event-by-event dynamic fluctuations lead to inconsistent results for different traditional methods, especially in small systems [33,34]. These weaken the sensitivity of collective measurements to quantitatively extract the properties of QGP. In the proposed idea, the reaction plane is directly determined by the initial geometry, independent of the final anisotropy from medium evolution, which gives natural resistance to the nonflow correlation from hadronic collisions. Furthermore, in the Good-Walker picture [35,36], the coherent photoproduction process can be viewed as production averaging over all possible nuclear configurations, which means that there is no event by event fluctuation. Therefore, the proposed approach can directly link the collective observables to the initial geometry and provide a perfect baseline to test the different methods of collectivity estimation experimentally. These are crucially important for understanding the collective and fluid-like phenomena, especially in small systems.

Inevitably, there are also limitations of the proposed approach. Due to the significant background production from hadronic interactions, the approach loses effectiveness toward central collisions. In comparison with traditional approaches, the trigger on the exclusive reaction would greatly reduce the statistics and potentially bias the impact parameter distributions. Furthermore, the resistance to event-by-event fluctuations makes the approach become invalid for the estimation of higher order flows (e.g., $v_3$) from fluctuations.

In summary, we demonstrate the idea of probing the reaction plane via the feature of linear polarization of the coherent photoproduction process in relativistic heavy-ion collisions and take the process of $\gamma + A \rightarrow \rho^0 + A \rightarrow \pi^+ + \pi^- + A$ to illustrate that. We estimated the resolution of the reaction plane determination from the proposed approach at typical RHIC and LHC collision energies, which possesses very little collision system and energy dependence. The proposed approach can directly link the collective observables to the initial geometry and provide a perfect baseline to test the traditional methods of collectivity estimation experimentally. There is no event-by-event fluctuation or nonflow contribution in the proposed method, which makes it cleaner and more powerful for understanding the collective and fluid-like phenomena, especially in small systems. In comparison with traditional approaches, despite the limitations, the proposed idea has unique advantages, not just a complementary method. The proposed idea is also applicable for other coherent photon produced vector mesons, e.g., $\omega$, $\phi$, $J/\psi$ and $\Upsilon$, which give similar results and conclusions.

This work is supported in part by the National Key Research and Development Program of China under Contract No. 2022YFA1604900 the National Natural Science Foundation of China (NSFC) under Contract No. 12175223 and 12005220. W.Z. is supported by Anhui Provincial Natural Science Foundation No. 2208085J23 and Youth Innovation Promotion Association of Chinese Academy of Sciences.

[1] Y. Aoki, G. Endrodi, Z. Fodor, S. D. Katz, and K. K. Szabo, The order of the quantum chromodynamics transition predicted by the standard model of particle physics, Nature (London) **443**, 675 (2006).

[2] P. Braun-Munzinger and J. Stachel, The quest for the quark-gluon plasma, Nature (London) **448**, 302 (2007).

[3] W. Busza, K. Rajagopal, and W. van der Schee, Heavy Ion Collisions: The Big Picture, and the Big Questions, Annu. Rev. Nucl. Part. Sci. **68**, 339 (2018).

[4] U. Heinz and R. Snellings, Collective flow and viscosity in relativistic heavy-ion collisions, Annu. Rev. Nucl. Part. Sci. **63**, 123 (2013).

[5] F. Becattini and M. A. Lisa, Polarization and Vorticity in the Quark–Gluon Plasma, Annu. Rev. Nucl. Part. Sci. **70**, 395 (2020).

[6] D. E. Kharzeev, J. Liao, S. A. Voloshin, and G. Wang, Chiral magnetic and vortical effects in high-energy nuclear collisions—A status report, Prog. Part. Nucl. Phys. **88**, 1 (2016).

[7] T. Hirano and M. Gyulassy, Perfect fluidity of the quark gluon plasma core as seen through its dissipative hadronic corona, Nucl. Phys. A **769**, 71 (2006).

[8] L. Adamczyk *et al.* (STAR), Global Λ hyperon polarization in nuclear collisions: evidence for the most vortical fluid, Nature (London) **548**, 62 (2017).

[9] X.-G. Huang, Electromagnetic fields and anomalous transports in heavy-ion collisions — A pedagogical review, Rep. Prog. Phys. **79**, 076302 (2016).

[10] K. Hattori and X.-G. Huang, Novel quantum phenomena induced by strong magnetic fields in heavy-ion collisions, Nucl. Sci. Tech. **28**, 26 (2017).

[11] J. Jia, Event-shape fluctuations and flow correlations in ultrarelativistic heavy-ion collisions, J. Phys. G: Nucl. Part. Phys. **41**, 124003 (2014).

[12] F. Krauss, M. Greiner, and G. Soff, Photon and gluon induced processes in relativistic heavy ion collisions, Prog. Part. Nucl. Phys. **39**, 503 (1997).

[13] C. A. Bertulani, S. R. Klein, and J. Nystrand, Physics of ultraperipheral nuclear collisions, Annu. Rev. Nucl. Part. Sci. **55**, 271 (2005).

[14] C. Li, J. Zhou, and Y.-J. Zhou, Probing the linear polarization of photons in ultraperipheral heavy ion collisions, Phys. Lett. B **795**, 576 (2019).

[15] J. Adam *et al.* (STAR), Measurement of $e^+e^-$ Momentum and Angular Distributions from Linearly Polarized Photon Collisions, Phys. Rev. Lett. **127**, 052302 (2021).

[16] L. Criegee *et al.*, ρ production with polarized photons, Phys. Lett. B **28**, 282 (1968).

[17] J. Ballam *et al.*, Study of $\gamma p \to p\omega$ with Linearly Polarized Photons at 2.8 and 4.7 GeV, Phys. Rev. Lett. **24**, 1364 (1970).

[18] Y. Eisenberg, B. Haber, E. Kogan, U. Karshon, E. E. Ronat, A. Shapira, and G. Yekutieli, Vector meson production by 4.3 GeV polarized photons on deuterium, Nucl. Phys. B **104**, 61 (1976).

[19] M. Derrick *et al.* (ZEUS), Measurement of elastic φ photoproduction at HERA, Phys. Lett. B **377**, 259 (1996).

[20] B. I. Abelev *et al.* (STAR), $\rho^0$ photoproduction in ultraperipheral relativistic heavy ion collisions at $\sqrt{s_{NN}} = 200$ GeV, Phys. Rev. C **77**, 034910 (2008).

[21] T. Hiraiwa, M. Yosoi, M. Niiyama, Y. Morino, Y. Nakatsugawa, M. Sumihama, D. S. Ahn, J. K. Ahn, W. C. Chang, J. Y. Chen, S. Date, H. Fujimura, S. Fukui, K. Hicks, T. Hotta, S. H. Hwang, T. Ishikawa, Y. Kato, H. Kawai, H. Kohri, Y. Kon, P. J. Lin, Y. Maeda, M. Miyabe, K. Mizutani, N. Muramatsu, T. Nakano, Y. Nozawa, Y. Ohashi, T. Ohta, M. Oka, C. Rangacharyulu, S. Y. Ryu, T. Saito, T. Sawada, H. Shimizu, E. A. Strokovsky, Y. Sugaya, K. Suzuki, A. O. Tokiyasu, T. Tomioka, T. Tsunemi, M. Uchida, T. Yorita, (LEPS), First measurement of coherent φ-meson photoproduction from $^4$He near threshold, Phys. Rev. C **97**, 035208 (2018).

[22] K. Schilling, P. Seyboth, and G. E. Wolf, On the Analysis of Vector Meson Production by Polarized Photons, Nucl. Phys. B **15**, 397 (1970), [Erratum: **18**, 332 (1970)].

[23] W. Zha, S. R. Klein, R. Ma, L. Ruan, T. Todoroki, Z. Tang, Z. Xu, C. Yang, Q. Yang, and S. Yang, Coherent J/ψ photoproduction in hadronic heavy-ion collisions, Phys. Rev. C **97**, 044910 (2018).

[24] R. C. Barrett, D. F. Jackson, and G. W. Greenless, Nuclear sizes and structure, Phys. Today **31**(7), 47 (1978).

[25] M. L. Miller, K. Reygers, S. J. Sanders, and P. Steinberg, Glauber modeling in high energy nuclear collisions, Annu. Rev. Nucl. Part. Sci. **57**, 205 (2007).

[26] T. H. Bauer, R. D. Spital, D. R. Yennie, and F. M. Pipkin, The hadronic properties of the photon in high-energy interactions, Rev. Mod. Phys. **50**, 261 (1978), [Erratum: **51**, 407 (1979)].

[27] S. R. Klein, J. Nystrand, J. Seger, Y. Gorbunov, and J. Butterworth, STARlight: A Monte Carlo simulation program for ultraperipheral collisions of relativistic ions, Comput. Phys. Commun. **212**, 258 (2017).

[28] J. Hüfner and B. Z. Kopeliovich, J/Psi N and Psi-prime N total cross-sections from photoproduction data: Failure of vector dominance, Phys. Lett. B **426**, 154 (1998).

[29] J. Adams *et al.* (STAR), Azimuthal anisotropy in Au+Au collisions at s(NN)**(1/2) = 200-GeV, Phys. Rev. C **72**, 014904 (2005).

[30] B. Schenke, P. Tribedy, and R. Venugopalan, Fluctuating Glasma Initial Conditions and Flow in Heavy Ion Collisions, Phys. Rev. Lett. **108**, 252301 (2012).

[31] A. Jaiswal and V. Roy, Relativistic hydrodynamics in heavy-ion collisions: general aspects and recent developments, Adv. High Energy Phys. **2016**, 9623034 (2016).

[32] J. Xu, Transport approaches for the description of intermediate-energy heavy-ion collisions, Prog. Part. Nucl. Phys. **106**, 312 (2019).

[33] K. Dusling, W. Li, and B. Schenke, Novel collective phenomena in high-energy proton–proton and proton–nucleus collisions, Int. J. Mod. Phys. E **25**, 1630002 (2016).

[34] J. L. Nagle and W. A. Zajc, Small system collectivity in relativistic hadronic and nuclear collisions, Annu. Rev. Nucl. Part. Sci. **68**, 211 (2018).

[35] M. L. Good and W. D. Walker, Diffraction disssociation of beam particles, Phys. Rev. **120**, 1857 (1960).

[36] T. Lappi and H. Mantysaari, Incoherent diffractive J/Psi-production in high energy nuclear DIS, Phys. Rev. C **83**, 065202 (2011).